\documentclass[11pt,aps,preprint,nofootinbib]{revtex4}
\usepackage[active]{srcltx}
\usepackage[utf8]{inputenc}
\usepackage{latexsym}
\usepackage{amsmath}
\usepackage{amssymb}
\usepackage{graphicx}
\usepackage{slashed}
\usepackage{xcolor}

\begin{document}

\title{Duality between 3D Massive Thirring and Maxwell Chern-Simons models from 2D bosonization}

\author{Carlos A. Hernaski}
\email{hernaski@uel.br}
\affiliation{Departamento de F\'isica, Universidade Estadual de Londrina, \\
Caixa Postal 10011, 86057-970, Londrina, PR, Brasil}

\author{Pedro R. S. Gomes}
\email{pedrogomes@uel.br}
\affiliation{Departamento de F\'isica, Universidade Estadual de Londrina, \\
Caixa Postal 10011, 86057-970, Londrina, PR, Brasil}

\begin{abstract}

Bosonization techniques are important nonperturbative tools in quantum field theory. In three dimensions they possess interesting connections to topologically ordered systems and ultimately have driven the observation of an impressive web of dualities.
In this work, we use the quantum wires formalism to show how the fermion-boson mapping relating the low-energy regime of the massive Thirring model in three spacetime dimensions with the Maxwell-Chern-Simons model can be obtained from the exact bosonization in two dimensions.

\end{abstract}
\maketitle


{\textbf{\textit{Introduction}}}. Bosonization is one of the main tools to analyze nonperturbative properties of quantum field theories and condensed matter systems in 1+1 dimensions (2D). From the properties of the Fermi surfaces in one spatial dimension, one can show that such a fermion-boson relation is always possible through an explicit operator identification of the fermion with the boson field \cite{Mandelstam}. This strict one-to-one correspondence between fermion and boson models, in general, does not survive in higher dimensions \cite{Haldane}. However, we can still find many instances where fermionic and bosonic models are dual within certain regimes in dimensions higher than two. Particularly, in 2+1 dimensions (3D) a conjectured fermion-boson relation has driven the observation of an impressive web of dualities, which has interesting connections to topologically ordered systems \cite{Karch,Witten,Tong,Nastase,Aharony}. Coupling with a Chern-Simons gauge field seems to play a decisive role in the promotion of bosonization to three dimensions. This goes back to the discussion in \cite{Polyakov}, which demonstrated the spin transmutation mechanism ruled by the Chern-Simons field when coupled to fermion and boson fields. 

The Thirring model (TM) has been an interesting arena of fruitful insights into the bosonization program since the works of \cite{Klaiber,Coleman,Mandelstam} in 2D. Using functional techniques, the authors in \cite{Fradkin} showed the relation between the massive TM in 3D with the Maxwell-Chern-Simons (MCS) model in the large mass limit. The emergence of a gauge invariance from the TM comes as a surprise in the approach of \cite{Fradkin}. In an independent line of development this matter is clarified starting with a reformulation of the TM as a gauge theory \cite{Gomes,Itoh,Kondo,Quevedo}. Furthermore, the idea is generalized to arbitrary dimensions and the connection between several intermediate models is revealed using master Lagrangians \cite{Ikegami,Burgess}. In addition to the interest in the context of high-energy physics, the TM model appears frequently in condensed matter systems. It plays an important role in the description of interacting electrons restricted to move in one spatial dimension (Luttinger model), where the bosonization is specially useful once it essentially provides an exact solution of the interacting problem \cite{Fradkin1}. The TM can also be relevant in higher-dimensional systems. In this line, the authors in \cite{Palumbo} discuss how the massive 3D TM can emerge from the low-energy limit of a tight-binding model of spinless fermions on a honeycomb lattice generalizing in this way the TM-MCS relation to condensed matter physics.

Our goal in this letter is to show how the fermion-boson map in 3D can be obtained from the exact bosonization in 2D. Contrary to the previous discussions we will use an operator-based approach to bosonization. Concretely, we split one of the spatial dimensions of the system and treat it as a discretized one, transforming the system into a set of quantum wires \cite{Kane,Teo,Neupert}. In this way we can use the bosonization recipe of 2D in each one of the wires. A similar strategy was used in \cite{Mross1} to derive a three-dimensional fermion-fermion duality presented in \cite{Son}. With this approach we are able to recover the MCS low-energy limit of the massive Thirring model and the bosonization rule for the fermionic current. We believe that this program can be useful for investigating further connections in the web of dualities in 3D.

{\textbf{\textit{Thirring model and quantum wires}}}. We start by considering the massive Thirring model in 2+1 dimensions
\begin{eqnarray}
\mathcal{L}&=&\bar{\psi}\mathrm{i}\slashed\partial\psi+M\bar{\psi}\psi-\frac{g^2}{2}J_\mu J^\mu\label{1},
\end{eqnarray}
with $J^\mu=\bar{\psi}\gamma^\mu\psi$. As is well known, this model is perturbatively non-renormalizable\footnote{On the other hand, in the case  of $N$ fermionic fields it is renormalizable in the large-$N$ expansion.}. To make sense of it we suppose the existence of a mass scale $\Lambda$ that cut off the high-energy modes of the spinor field. Using this mass scale, we can define the dimensionless coupling constant $\lambda^2\equiv g^2\Lambda$. In the weak coupling perturbative regime the current-current interaction is irrelevant and should play no role in the low-energy regime of the model. However, we are just interested in a possible strong coupling scenario, $\lambda^2\gg 1$, where a perturbative treatment is not applicable.
Using functional methods, the authors in \cite{Fradkin} show that the model (\ref{1}) is equivalent to the Maxwell-Chern-Simons theory in the infinity mass limit $M\rightarrow\infty$. It is clear that in the low energy regime, $E\ll \Lambda$, the particle states created by the fermion operator cannot be excited in this limit, and the resulting nontrivial low-energy theory signals for the existence of fermion-antifermion bound states in the low-energy spectrum.

We will use the 2D operator formalism to understand how the 3D bosonization can follow from the fermion-boson operator relation in 2D. To this end, we consider a set of quantum wires with Hamiltonian density
\begin{eqnarray}
\mathcal{H}&=&\sum_{i=1}^N-\psi^{i\dagger}_R\mathrm{i}\partial_x\psi^i_R+\psi^{i\dagger}_L\mathrm{i}\partial_x\psi^i_L+\frac{1}{a}\left(\psi^{i+1\dagger}_L\psi^i_R-\psi^{i\dagger}_L\psi^i_R+\text{H. c.}\right)\nonumber\\
&&+M\left(\psi^{i\dagger}_L\psi^{i}_R+\text{H. c.}\right)+\frac{\lambda^2_a}{2}\left(\left(J^{i}_L\right)^2+\left(J^i_R\right)^2\right)+\lambda^2_bJ^i_LJ^i_R-\lambda^2_c\left(J^i_y\right)^2,\label{2}
\end{eqnarray}
where $a$ is a short-distance cutoff, which can be understood as the interwire spacing. This Hamiltonian follows from the discretization of the $y$ direction in the Lagrangian (\ref{1}) and the identification $\psi(t,x,y)\rightarrow\frac{1}{\sqrt{a}}\psi^i(t,x)$. We are using the conventions: $\gamma^0=\sigma^1$, $\gamma^1=-\mathrm{i}\sigma^2$, $\gamma^2=\mathrm{i}\sigma^3$, and $\eta^{\mu\nu}=(1,-1,-1)$, with the two-component spinor $\psi^T=(\psi_R,\psi_L)$. The fields $\psi^i_{L/R}(t,x)$ are then interpreted as left and right moving fermions within each quantum wire. Upon discretization, the derivative term in the $y$ direction in (\ref{1}) is replaced by the hopping term in the first line. We also have included the $\left(J_{L/R}\right)^2$ terms compared to the original Lagrangian. The effect of these terms is only to renormalize the velocity of the chiral modes. We need such a possible velocity renormalization since we are starting our analysis from the Hamiltonian instead of the Lagrangian. Since the Hamiltonian is not Lorentz invariant, the regularization used to define it properly will give rise to Lorentz noninvariant renormalizations. An example is the distance cutoff $a$, which breaks Lorentz symmetry, contrary to the Lorentz invariant mass cutoff $\Lambda$ introduced in (\ref{1}). For the same reason we left the coupling constants $\lambda_a$, $\lambda_b$, and $\lambda_c$, completely unrelated at this point. After we take care of the divergences present in (\ref{2}), bosonize and take the continuum limit again, we will adjust the renormalization conditions to render Lorentz invariant quantities.

The original model (\ref{1}) has a $U(1)$ global symmetry $\psi^{\prime}=e^{\mathrm{i}\alpha}\psi$, which leads to the current conservation law $\partial_\mu J^\mu=0$. From the coupled set of quantum wires in (\ref{2}), this global symmetry gives the conservation of the total current $\sum_iJ^i_\alpha$, with $\alpha=t,x$. To track back the local charge dynamics in the quantum wires system we need to investigate the flow of charge into and out of each wire due to charge tunneling interaction operators, i.e., the operators that break the $U(1)$ symmetry $\psi^{i\prime}=e^{\mathrm{i}\alpha^i}\psi^i$, which 
corresponds to an independent global transformation for each one of the wires. Pursuing this analysis, one obtains the expression of the intrawire charge nonconservation $\partial_tJ^{it}+\partial_x J^{ix}=\frac{\mathrm{i}}{a}\left(\psi^{i+1\dagger}_L\psi^i_R-\psi^{i\dagger}_L\psi^{i-1}_R-\text{H. c.}\right)\equiv \Delta_i\left(\mathrm{i}\psi^{i+1\dagger}_L\psi^i_R+\text{H. c.}\right)$, with $J^{it}=\bar{\psi}^i\gamma^{0}\psi^i$, $J^{ix}=\bar{\psi}^i\gamma^{1}\psi^i$  and $\Delta_i$ is a discretized derivative operator. This equation recovers the information of the local charge conservation in the quantum wires system. With this discussion we identify the currents 
\begin{eqnarray}
J^i_{L/R}=\psi^{i\dagger}_{L/R}\psi^i_{L/R}~~~\text{and}~~~
J^i_y=\mathrm{i}\psi^{i+1\dagger}_L\psi^i_R+\text{H. c.}\label{4}
\end{eqnarray}
in the interactions in (\ref{2}). As usual, these currents should be defined with the usual normal ordering point splitting prescription.

{\textbf{\textit{Bosonization}}.} Following the conventions of \cite{Teo}, we bosonize the Hamiltonian (\ref{2}) according to the fermion-boson mapping:
\begin{eqnarray}
\psi^i_p&=&\frac{\kappa^i}{\sqrt{2a\pi}}e^{\mathrm{i}\left(\varphi^i+p\vartheta^i\right)}\label{5},
\end{eqnarray}
with $p=R/L=+1/-1$. The boson field $\varphi$ and its dual $\vartheta$ are defined in terms of the chiral boson fields $\phi_{L/R}$ as $\varphi^i\equiv \left(\phi^i_R+\phi^i_L+\pi N^i_L\right)/2$ and $\vartheta^i\equiv \left(\phi^i_R-\phi^i_L+\pi N^i_L\right)/2$. The $\kappa^i$ are Klein factors given in terms of the number operators $N^i_p=\frac{p}{2\pi}\int{dx\partial_x\phi^i_p}$ according to $\kappa^i=\left(-1\right)^{\sum_{j<i}N_L^j+N_R^j}$. For our purposes the important commutation relations are $\left[\vartheta^i(x),\varphi^j(x^\prime)\right]=\mathrm{i}\pi\delta_{ij}\Theta(x-x^\prime)$, $\left[\vartheta^i(x),\vartheta^j(x^\prime)\right]=\left[\varphi^i(x),\varphi^j(x^\prime)\right]=0$, and $\left[N_p^i,\phi_q^j(x)\right]=\mathrm{i}\delta_{ij}\delta_{pq}$. The fundamental excitations created by the bosonic field $\vartheta$ are fermion-antifermion bound states. Local polynomials of this field only span the null charge sector of the model. To create a charged state inside a wire one needs extended soliton solutions. So, the fermion can be seen as coherent bound state, as expressed by (\ref{5}). This physical interpretation can be inferred by noticing that the charge density operator can be shown to be given by $\rho^i(x)=\partial_x\vartheta^i(x)/\pi$. A unit of charge within the wire then occurs when $\vartheta$ has a kink where it jumps by $\pi$. Nontrivial topological sectors within the bosonic theory are accounted for by the number operators $N^i_p$, which essentially count the solitons inside a wire.

A direct application of the above rules in the currents in (\ref{4}) as well as in the other terms of (\ref{2}) gives us the bosonic Hamiltonian density
\begin{eqnarray}
\mathcal{H}&=&\sum_{i=1}^N\frac{v}{2\pi}\left(K\left(\partial_x\varphi^{i}\right)^2+\frac{1}{K}\left(\partial_x\vartheta^{i}\right)^2\right)+\frac{1}{a\pi}\left(M-\frac{1}{a}\right)\sin\left(2\vartheta^{i}\right)\nonumber\\
&&-\frac{\lambda^2_c}{\pi^2 a^2}\left(\cos\left(\varphi^{i+1}-\varphi^i-\vartheta^{i+1}-\vartheta^i+\pi N^i\right)\right)^2-\frac{1}{a^2\pi}\sin\left(\varphi^{i+1}-\varphi^i-\vartheta^{i+1}-\vartheta^i+\pi N^i\right),\label{6}
\end{eqnarray}
with $N^i=N^i_L+N^i_R$, $v=\sqrt{\left(1+\frac{\lambda^2_a}{2\pi}\right)^2-\left(\frac{\lambda^2_b}{2\pi}\right)^2}$, and $K=\sqrt{\left(1+\frac{\lambda^2_a}{2\pi}-\frac{\lambda^2_b}{2\pi}\right)/\left(1+\frac{\lambda^2_a}{2\pi}+\frac{\lambda^2_b}{2\pi}\right)}$.
The quadratic Hamiltonian describes a gapless system in a sliding Luttinger liquid phase, whereas the sine and cosine operators can destabilize the phase and open a mass gap in the system.

\textbf{\textit{Continuum limit}.} In order to safely take the low-energy continuum limit of (\ref{6}), it is convenient to remove the divergences of the model by normal ordering the operators. Normal ordering the quadratic terms only gives an additive renormalization of the zero point energy. For the interaction operators, we need to use a basic rule for the exponential of operators $e^Ae^B=e^{A+B}e^{[A,B]/2}$ for a c-number commutator $[A,B]$. Then, we can show that for local fields $A$ and $B$ we have $e^{A}=:e^{A}:e^{\frac{1}{2}\left<AA\right>}$, $:e^{A}::e^{B}:=:e^{A+B}:e^{\left<AB\right>}$, with $:e^{A}:\equiv e^{A^+}e^{A^-}$, and $A^+$ and $A^-$ being the creation and annihilation parts of the field $A$. Using the equal time correlations $\left<\varphi^{i}(x^\prime)\varphi^j(x)\right>=\left<\vartheta^{i}(x^\prime)\vartheta^j(x)\right>=-\frac{1}{4}\ln\left(\mu^2\left(\left(x^\prime-x\right)^2+a^2\right)\right)\delta^{ij}$, where $\mu$ is an infrared mass that will be fixed soon, to normal order the Hamiltonian and making the rescalings $\varphi^i\rightarrow\varphi^i/\sqrt{K}$ and $\vartheta^i\rightarrow\sqrt{K}\vartheta^i$, we get
\begin{eqnarray}
\mathcal{H}&=&\sum_{i=1}^N\frac{v}{2\pi}\left(:\left(\partial_x\varphi^{i}\right)^2:+:\left(\partial_x\vartheta^{i}\right)^2:\right)+\frac{\mu}{\pi}\left(M-\frac{1}{a}\right):\sin\left(2\sqrt{K}\vartheta^{i}\right):\nonumber\\
&&-\frac{\lambda^2_c\mu^2}{\pi^2}\left(:\cos\left(\frac{1}{\sqrt{K}}\left(\varphi^{i+1}-\varphi^i\right)-\sqrt{K}\left(\vartheta^{i+1}+\vartheta^i+\pi N^i\right)\right):\right)^2\nonumber\\
&&-\frac{\mu}{a\pi}:\sin\left(\frac{1}{\sqrt{K}}\left(\varphi^{i+1}-\varphi^i\right)-\sqrt{K}\left(\vartheta^{i+1}+\vartheta^i+\pi N^i\right)\right):.\label{7}
\end{eqnarray}

By expanding the interaction operators we would find linear terms in the fields. This suggests a redefinition of the vacuum of the theory as $\left<\vartheta^i\right>=-\frac{\pi}{4\sqrt{K}}$, which gives
\begin{eqnarray}
\mathcal{H}&=&\sum_{i=1}^N\frac{v}{2\pi}\left(:\left(\partial_x\varphi^{i}\right)^2:+:\left(\partial_x\vartheta^{i}\right)^2:\right)-\frac{\mu}{\pi}\left(M-\frac{1}{a}\right):\cos\left(2\sqrt{K}\vartheta^{i}\right):\nonumber\\
&&-\frac{\lambda^2_c\mu^2}{\pi^2}\left(:\sin\left(\frac{1}{\sqrt{K}}\left(\varphi^{i+1}-\varphi^i\right)-\sqrt{K}\left(\vartheta^{i+1}+\vartheta^i+\pi N^i\right)\right):\right)^2\nonumber\\
&&-\frac{\mu}{a\pi}:\cos\left(\frac{1}{\sqrt{K}}\left(\varphi^{i+1}-\varphi^i\right)-\sqrt{K}\left(\vartheta^{i+1}+\vartheta^i+\pi N^i\right)\right):.\label{a1}
\end{eqnarray}

To take the continuum limit, we rescale the bosonic fields as $\Sigma(t,x)\rightarrow\sqrt{a}\Sigma(t,x,y)$, which implies $\Sigma^{i+1}\rightarrow\sqrt{a}\Sigma+a^{3/2}\partial_y\Sigma$ up to irrelevant higher derivative terms. Here, $\Sigma$ stands for either $\varphi$ or $\vartheta$. The Hamiltonian (\ref{a1}) can then be expanded as
\begin{eqnarray}
\mathcal{H}&=&\int{dy\frac{v}{2\pi}\left(\left(\partial_x\varphi\right)^2+\left(\partial_x\vartheta\right)^2\right)-\frac{\mu}{a\pi}M-\frac{\mu}{a\pi}\left(M-\frac{1}{a}\right)2Ka\vartheta^2}\nonumber\\
&&+\left(\frac{1}{2a}-\frac{\lambda^2_c\mu}{\pi}\right)\frac{\mu a^2}{\pi K}\left(\partial_y\varphi-\frac{K}{a}\left(2\vartheta+a\partial_y\vartheta+\pi N\right)\right)^2+\ldots,\label{a2}
\end{eqnarray}
where the dots represent higher power of the fields. To get the large-$M$ limit, we identify $M$ with the inverse cutoff $1/a$. Then, from the expression above we can read the squared mass $4\left(\frac{1}{2}-\frac{g^2_c\mu}{\pi}\right)K\mu M$ of the $\vartheta$ field, with $g^2_c=a\lambda^2_c$. According with our initial definition of the two-point function of this field, $\mu^2$ should be identified with this mass. This is essentially a renormalization condition for the two-point function. This consistency identification gives $K=\alpha\mu/M$, with $\alpha=\frac{\pi}{4\left(\pi-2g_c^2\mu\right)}$ being a finite number. The important point is that $K$ goes like $\mu/M$ and then we get the Hamiltonian
\begin{eqnarray}
H&=&\frac{1}{2\pi}\int{d^2\vec{r}\left[v\left(\left(\partial_x\varphi\right)^2+\left(\partial_x\vartheta\right)^2\right)+\frac{1}{4\alpha^2}\left(\partial_y\varphi-2\mu\alpha\left(\vartheta+\frac{\pi}{2}N\right)\right)^2\right]}+\mathcal{O}(1/M).\label{a3}
\end{eqnarray}
where we have omitted the normal ordering symbol for the quadratic terms.

\textbf{\textit{Maxwell-Chern-Simons}.} The crucial step to show the equivalence with Maxwell-Chern-Simons theory is the identification of the components of electromagnetic field as
\begin{equation}
B=\sqrt{\frac{v}{\pi}}\partial_x\vartheta, ~~~E^{y}=\sqrt{\frac{v}{\pi}}\partial_x\varphi~~~\text{and}~~~ E^{x}=-\frac{1}{\sqrt{4\alpha^2\pi}}\left(\partial_y\varphi-2\alpha\mu\left(\vartheta+\frac{\pi}{2}N\right)\right).
\end{equation}
This is how the "microscopic" 1+1 dimensional variables are related to the  "macroscopic" 2+1 dimensional fields. Notice that this is an identification between physical fields, i.e., between gauge invariant quantities. With this and taking the limit $M\rightarrow\infty$ we can put the Hamiltonian (\ref{a3}) into the form
\begin{eqnarray}
H&=&\frac{1}{2}\int{d^2\vec{r}\left(\vec{E}^2+B^2\right)},\label{9}
\end{eqnarray}
which is formally identical to the Maxwell-Chern-Simons Hamiltonian written in terms of the electric and magnetic fields $\vec{E}$ and $B$. However, this identification alone is not enough to ensure the equivalence. We need additionally to show that these fields defined in terms of $\varphi$ and $\vartheta$ satisfy the algebra \cite{Deser}
\begin{eqnarray}
\left[E^{a}(\vec{r}),B(\vec{r}^\prime)\right]=\mathrm{i}\epsilon^{ab}\partial_b\delta(\vec{r}-\vec{r}^\prime)~~~\text{and}~~~
\left[E^{a}(\vec{r}),E^{b}(\vec{r}^\prime)\right]=-\mathrm{i}\mu\epsilon^{ab}\delta(\vec{r}-\vec{r}^\prime),\label{10}
\end{eqnarray}
where the indexes $a,b$ correspond to the $x,y$ components of the fields.
Using the commutators for the $\varphi^i$ and $\vartheta^i$ fields, we can show this is in fact the case provided the conditions $v=1$ and $\alpha=1/2$ are imposed on the parameters of the Hamiltonian. These conditions are the adjustment in the finite renormalizations to match the relativistic dynamics. Thus out of four initial parameters,      $\lambda_a^2=g_a^2/a$, $\lambda_b^2=g_b^2/a$, $\lambda_c^2=g_c^2/a$, and $\mu$, the two conditions $v=1$ and $\alpha=1/2$ together with the renormalization condition for the two-point function of $\vartheta$ ($K=\alpha \mu /M$), leave us with only one independent constant, say $ \lambda_c^2=g_c^2/a$. This is precisely the number of independent parameters of the large mass limit of the 3D Thirring model and also of the MCS model.
The explicit solutions to the above conditions are $\mu=\pi/4g_c^2$, $\lambda_b^2=8\lambda_c^2\left[1-(\pi/8\lambda_c^2)^2\right]$, and 
$\lambda_a^2=2\pi\left[\sqrt{1+\left[\frac{4\lambda_c^2}{\pi}\left(1-(\frac{\pi}{8\lambda_c^2})^2\right)\right]^2}-1\right]$. With this, we note in particular that the algebra (\ref{10}) can be then written in terms of the coupling constant $g_c^2$. This coupling constant, in turn, can be related to the macroscopic coupling constant $g^2$ of the 3D Thirring interaction through $g_c^2=g^2/8$. The algebra (\ref{10}) corresponds to the Maxwell-Chern-Simons algebra that follows from the Lagrangian
\begin{equation}
\mathcal{L}=-\frac{1}{4}F_{\mu\nu}F^{\mu\nu}+\frac{\pi}{g^2}\epsilon^{\mu\nu\rho}A_\mu\partial_\nu A_\rho.\label{e1}
\end{equation}

\textbf{\textit{Current Algebra}.} From the 2D bosonization relations we have identified the electromagnetic fields in terms of the bosonic fields in the large mass limit. Let us investigate the large mass limit directly in the components of the Thirring current $\tilde{J}^\mu=\bar{\tilde{\psi}}\gamma^\mu\tilde{\psi}$, where $\tilde{\psi}$ stands for the Fermion field (\ref{5}) with the rescaled bosonic fields $\frac{1}{\sqrt{K}}\varphi$ and $\sqrt{K}\vartheta$. It is then straightforward to obtain the bosonized components $\tilde{J}^i_0=\tilde{J}^i_R+\tilde{J}^i_L=\frac{\sqrt{K}}{\pi}\partial_x\vartheta^i$ and $\tilde{J}^{xi}=\tilde{J}^i_R-\tilde{J}^i_L=\frac{1}{\pi\sqrt{K}}\partial_x\varphi^i$. Similarly, for the component 
$\tilde{J}^{yi}=\mathrm{i}\tilde{\psi}^{i+1\dagger}_L\tilde{\psi}^i_R+\text{H. c.}$ we first obtain
\begin{equation}
\tilde{J}^{yi}=-\frac{1}{\pi a}\cos\left(\frac{1}{\sqrt{K}}\left(\varphi^{i+1}-\varphi^i\right)-\sqrt{K}\left(\vartheta^{i+1}+\vartheta^i+\pi N^i\right)\right). 
\end{equation}
As before, we identify $K=\pi/Mg^2$, make the shift $\vartheta^i\rightarrow\vartheta^i-\frac{\pi}{4\sqrt{K}}$, and to take the large mass limit we also make $a=1/M$. Then, in the limit $M\rightarrow\infty$ we get $\tilde{J}^{yi}=\frac{1}{\pi\sqrt{K}}\left(\partial_y\varphi^i-\frac{2\pi}{g^2}\left(\vartheta^i+\frac{\pi}{2}N^i\right)\right)$. Because of the difference in place of the normalization factor $\sqrt{K}$ in the $\tilde{J}^i_0$ component compared to $\tilde{J}^{xi}$ and $\tilde{J}^{yi}$, these three components do not form a covariant three-vector. This situation is similar to that one found in the Thirring current in 2D. It is known from \cite{Klaiber} that an extra $K$ factor is needed in the definition of the spatial component of the current for a correct treatment of the infinities. Analogously, we then redefine the current as $\tilde{J}^{i\mu}=\left(\delta^\mu_0+K\delta^\mu_{x,y}\right)\bar{\tilde{\psi}}^i\gamma^\mu\tilde{\psi}^i$, with no sum over $\mu$. With the identification of the electromagnetic fields above, we then obtain the usual bosonization rule for the Thirring current in the large mass limit
$J^{\mu i}_{Th}=\sqrt{\frac{K}{\pi}}\epsilon^{\mu\nu\rho}\partial_\nu A^{i}_{\rho}$,
with $A^i_\mu$ being the potential three-vector, such that $B^i=\partial_xA^i_y-\partial_yA^i_x$, $E^i_x=\partial_tA^i_x-\partial_xA^i_t$, and $E^i_y=\partial_tA^i_y-\partial_yA^i_t$. The 2+1 Thirring current $J^\mu$ is obtained from $J^{i\mu}$ by just a rescaling, $J^\mu=\frac{1}{a}J^{i\mu}$, which gives
\begin{equation}
J^\mu_{Th}=\sqrt{\frac{1}{ g^2}}\epsilon^{\mu\nu\rho}\partial_\nu A_{\rho},
\label{12}
\end{equation}
with $A_\mu=\frac{1}{\sqrt{a}}A^i_\mu$. In particular, 
\begin{eqnarray}
\left[J^0_{Th}(t,\vec{x}),J^a_{Th}(t,\vec{y})\right]&=&\mathrm{i}\frac{1}{g^2}\epsilon^{ab}\partial_b\delta(\vec{x}-\vec{y}),\label{13}\\
\left[J^a_{Th}(t,\vec{x}),J^b_{Th}(t,\vec{y})\right]&=&-\mathrm{i}\frac{1}{g^4}\epsilon^{ab}\delta(\vec{x}-\vec{y}),\label{14}
\end{eqnarray}
which is finite in the limit $M\rightarrow\infty$ in agreement with the discussion in \cite{Schaposnik}.
It is important to mention that one should be careful in taking the limit $g\rightarrow 0$ in our expressions, since this would imply a weak coupling regime of the model and,  consequently, a breaking of our starting assumption. In fact, one can verify that this leads to singular expressions in many places. However, it is interesting to notice that if this limit is taken in the algebra (\ref{13}) and (\ref{14}) one recovers the infinite Schwinger term for free fermions.

At this point it is interesting to discuss how the symmetries are matched on both sides of the duality. The discrete symmetries are easily compared. Because of the mass term,  the Thirring model is not invariant under the inversions\footnote{The inversions are defined by the operation of reversing the sign of one of the spacetime coordinates.} $P$ and $T$, but preserves charge conjugation $C$ and the combination $PT$. The same is true for the bosonic model, since the CS term has the same properties as the fermion mass term while the Maxwell term preserves all the discrete symmetries \cite{Deser}. Concerning the continuous symmetries, the first puzzle is the presence of a $U(1)$ gauge symmetry in the bosonic model, which is not evident in the fermionic one. But since this is a local symmetry, the mismatch should not worry us. In fact it is long known that even the Thirring model can be turned into a gauge theory by introducing auxiliary fields \cite{Gomes,Quevedo,Itoh,Kondo,Burgess,Ikegami}. Of more relevance is the pairing of global symmetries, since these are connected with the existence of conserved charges. In the Thirring model there is a global $U(1)$ symmetry, which leads to the conservation of electric charge, whereas in the MCS model there is an exactly conserved topological current ($\sim \epsilon^{\mu\nu\rho}\partial_{\nu}A_{\rho}$), 
with the flux of the magnetic field being the associated charge. These are tied through the bosonization map (\ref{12}). We can check this correspondence: As we have discussed in the Thirring model, in addition to the charged states created by the fermion field, we also have, at the strong coupling limit, charge-zero bound states present in the spectrum. By focusing on the low-energy dynamics and taking the large fermion mass limit, only the bound states remain. According to the duality, this regime is mapped to the MCS model, which also describes only charge-zero states. In fact, considering the field equation $\partial_i F^{i0}+\frac{4\pi}{g^2}B=0$ and integrating over a spatial surface we obtain $\int{d^2x B}=0$, after discarding a surface integral of the electric field, since it decreases exponentially due to the massive character of the gauge fields.

\textbf{\textit{Conclusions}.} We have derived the 3D fermion-boson mapping relating the low-energy regime of the massive Thirring model with the Maxwell-Chern-Simons theory from the exact bosonization rules in 2D. This has been done by discretizing one spatial dimension and then proceeding with the operator formalism. 
A natural question concerns with the extension of the results to the case of finite mass charged sectors of the Thirring model. In this situation, the fermionic excitations should be captured on the bosonic side through the phenomenon of flux attachment, where a given field can transmute its statistics when coupled to a Chern-Simons gauge field. It is essential at this point that the Chern-Simons terms have properly quantized levels and the gauge field configurations have quantized magnetic fluxes. In this sense, it is natural to investigate if the Chern-Simons term in our discussion is the one responsible for statistics transmutation. If this is the case the Chern-Simons level should also be quantized despite the fact that we are in the zero charge sector of the model and the gauge field configurations have vanishing flux. 

To explore this possibility, it is convenient to bring the normalization of the current (\ref{12}) to the more usual one that generates the quantized fluxes according to $Q=\int{J^0d^2x}=\int{\frac{B}{2\pi}d^2x}=\mathbb{Z}$. To reach this normalization we need to rescale the gauge fields as $A_\mu\rightarrow\frac{\sqrt{g^2}}{2\pi}A_\mu$. This rescaling changes the Chern-Simons coefficient in (\ref{e1}) to $1/4\pi$, which corresponds to a Chern-Simons level $1$. With this level, attaching flux to a scalar field, for example, can change its statistics to a fermion. This is compatible with the scenario where the Chern-Simons term appearing in our study is, in fact, the one responsible for statistics transmutation and can be an important clue to the extension of the duality to the finite mass charged sectors of the Thirring model.

In addition to offering a new perspective on the bosonization program in higher dimensions, we believe the formalism discussed is general and useful to study a wider class of fermion-boson relations in 3D, playing an important role in the web of dualities. Finally, our construction can also be related with the description of topological phases of matter in terms of quantum wires, specifically, with the Abelian quantum Hall phases discussed in  \cite{Kane,Teo}. In this context, this work provides an interesting starting point for  the establishment of a concrete bridge between microscopic theories based on fermionic degrees of freedom and effective low-energy topological field theories given in terms of the Chern-Simons action.

\textbf{\textit{Acknowledgments}.} We wish to thank David Tong and Carlos Núñez for the enlightening discussions and the useful comments on the manuscript. We acknowledge the financial support of Brazilian agencies CAPES and CNPq.



\begin{thebibliography}{99}

\bibitem{Mandelstam} S. Mandelstam, {\it Soliton operators for the quantized sine-Gordon equation}, Phys. Rev. D11, 3026, (1975).

\bibitem{Haldane} F. D. M. Haldane {\it Luttinger liquid theory of one-dimensional quantum fluids: I. Properties of the Luttinger model and their extension to the general 1D interacting spinless Fermi gas}, J. Phys. C: Solid State Phys., 14,  2585-2609, (1981).

\bibitem{Karch} A. Karch and D. Tong, {\it Particle-Vortex Duality from 3d Bosonization}, Phys. Rev. X6, 031043, (2016), arXiv:1606.01893 [hep-th].

\bibitem{Witten} N. Seiberg, T. Senthil, C. Wang, and E. Witten, {\it A Duality Web in 2+1 Dimensions and Condensed Matter Physics}, Annals Phys. 374, 395–433, (2016),
arXiv:1606.01989 [hep-th].

\bibitem{Tong} A. Karch, B. Robinson, and D. Tong, {\it More Abelian Dualities in 2+1 Dimensions}, JHEP 01, 017, (2017), arXiv:1609.04012 [hep-th].

\bibitem{Nastase} H. Nastase and C. Nunez, {\it Deriving three-dimensional bosonization and the duality web}, Phys.Lett. B776, 145, (2018), arXiv:1703.08182.  

\bibitem{Aharony} O. Aharony, F. Benini, P.-S. Hsin, N. Seiberg, {\it Chern-Simons-matter dualities with $SO$ and $USp$ gauge groups}, JHEP 1702, 072, (2017), arXiv:1611.07874.

\bibitem{Polyakov} A. M. Polyakov, {\it Fermi-Bose transmuations induced by gauge fields}, Mod. Phys. Lett. A3, 325, (1988).

\bibitem{Klaiber} B. Klaiber, {\it Lectures in Theoretical Physics}, edited by A. O. Barut and W. E. Brittin (Gordon and Breach, New York, 1968).

\bibitem{Coleman} S. Coleman, {\it Quantum sine-Gordon equation as the massive Thirring model}, Phys. Rev. D11, 2088, (1975).

\bibitem{Fradkin} E. H. Fradkin and F. A. Schaposnik, {\it The Fermion-Boson
Mapping in Three-Dimensional Quantum Field Theory}, Phys. Lett. B 338, 253, (1994).

\bibitem{Gomes} M. Gomes, R. S. Mendes, R. F. Ribeiro and A. J. da Silva, {\it Gauge structure, anomalies, and mass generation in a three-dimensional Thirring model}, Phys. Rev. D43, 3516, (1991).

\bibitem{Quevedo} C. P. Burgess and F. Quevedo, {\it Bosonization as duality}, Nucl. Phys. B421, 373–390, (1994), arXiv:hep-th/9401105 [hep-th].

\bibitem{Itoh} T. Itoh, Y. Kim, M. Sugiura and K. Yamawaki, {\it Thirring Model as a Gauge Theory}, Prog. Theor. Phys. 93, 417, (1995).

\bibitem{Kondo} K.- I. Kondo, {\it Bosonization and Duality of Massive Thirring Model}, Prog. Theor. Phys. 94, 899, (1995).

\bibitem{Burgess} C.P. Burgess, C.A. Lutken, F. Quevedo, {\it Bosonization in higher dimensions}, Phys. Lett. B336, 18, (1994), hep-th/9407078.

\bibitem{Ikegami} K. Ikegami, K.- I. Kondo, A. Nakamura {\it Bosonization of Thirring Model in Arbitrary Dimension}, Prog. Theor. Phys. 95, 203, (1995).

\bibitem{Fradkin1} E. Fradkin, {\it Field Theories of Condensed Matter Systems}, Cambridge University Press, New York, 2013.

\bibitem{Palumbo} G. Palumbo and J. Pachos {\it Abelian Chern-Simons-Maxwell Theory from a Tight-Binding Model of Spinless Fermions}, Phys. Rev. Lett. 110, 211603, (2013).

\bibitem{Kane} C. L. Kane, R. Mukhopadhyay and T.C. Lubensky, {\it The fractional quantum Hall effect in an array of quantum wires}, Phys. Rev. Lett. 88, 036401, (2002), [cond-mat/0108445].

\bibitem{Teo} J. C. Teo and C. Kane, {\it From Luttinger liquid to non-Abelian quantum Hall states}, Phys. Rev. B 89, 085101 (2014), arXiv:1111.2617.

\bibitem{Neupert} T. Neupert, C. Chamon, C. Mudery, and R. Thomale, {\it Wire descontructioninsm of two-dimensional topological phases}, 
Phys. Rev. B, 205101, (2014), arXiv:1403.0953. 

\bibitem{Mross1} David F. Mross, Jason Alicea, Olexei I. Motrunich, {\it Explicit derivation of duality between a free Dirac cone and quantum electrodynamics in (2+1) dimensions}, Phys. Rev. Lett. 117, 016802 (2016), arXiv:1510.08455.

\bibitem{Son} D. T. Son, {\it Is the Composite Fermion a Dirac Particle?}, Phys. Rev. X5, 031027, (2015), arXiv:1502.03446.

\bibitem{Deser} S. Deser, R. Jackiw, and S. Templeton, {\it Topologically Massive Gauge Theories}, Annals Phys. 140, 372, (1982). 

\bibitem{Schaposnik} J. C. Le Guillou, C. Núñez, F. A. Schaposnik, {\it Current algebra and bosonization in three-dimensions}, Annals Phys. 251, 426-441, (1996).



\end{thebibliography}
\end{document}